\RequirePackage{etoolbox}
\csdef{input@path}{%
 {sty/}% cls, sty files
 {img/}% eps files
}%
\csgdef{bibdir}{bib/}% bst, bib files

\documentclass[ba]{imsart}
\pubyear{0000}
\volume{00}
\issue{0}
\doi{0000}
\firstpage{1}
\lastpage{1}

\usepackage{amsthm}
\usepackage{amsmath}
\usepackage{natbib}
\usepackage{hyperref}
\usepackage{graphicx}
\usepackage{graphicx}
\usepackage{amsmath}
\usepackage{times}
\usepackage{bm}
\usepackage{natbib}
\usepackage{amssymb}

\usepackage{color} 
\usepackage{changes}
\usepackage{array,bbm}

\usepackage{comment}
\usepackage{xr}
\usepackage{algorithm}% http://ctan.org/pkg/algorithms
\usepackage{algpseudocode}% http://ctan.org/pkg/algorithmicx
\usepackage{url}
\usepackage{soul}
\usepackage[textsize=tiny]{todonotes}    

\newcommand{\BB}[1]{\boldsymbol{#1}}

\newcommand{\R}{\mathbb R}

\def\beq{\begin{equation}}
\def\eeq{\end{equation}}
\def\bqn{\begin{eqnarray}}
\def\eqn{\end{eqnarray}}
\def\bqns{\begin{eqnarray*}}
\def\eqns{\end{eqnarray*}}
\def\bep{\begin{proof}}
\def\ep{\end{proof}}

\def\bc{\begin{center}}
\def\ec{\end{center}}

\def\s{\sigma}

\startlocaldefs
\numberwithin{equation}{section}
\theoremstyle{plain}
\newtheorem{thm}{Theorem}[section]
\endlocaldefs
\newcommand{\stkout}[1]{\ifmmode\text{\sout{\ensuremath{#1}}}\else\sout{#1}\fi}
\begin{document}
\date{}

\begin{frontmatter}
\title{A novel algorithmic approach to Bayesian Logic Regression\thanksref{T1}}
\runtitle{A novel algorithmic approach to Bayesian Logic Regression}
\thankstext{T1}{The first two authors gratefully acknowledge the financial support of the \textit{CELS project at the University of Oslo}, \url{http://www.mn.uio.no/math/english/research/groups/cels/index.html}.}

\begin{aug}
\author{\fnms{Aliaksandr} \snm{Hubin}\thanksref{addr1,addr4}},
\author{\fnms{Geir} \snm{Storvik}\thanksref{addr2}}
\and
\author{\fnms{Florian} \snm{Frommlet}\thanksref{addr3}}
\ead[label=e3]{florian.frommlet@meduniwien.ac.at}
\ead[label=e1]{aliaksah@math.uio.no}
\ead[label=e2]{geirs@math.uio.no}
\ead[label=e4]{aliaksandr.hubin@nr.no}

\runauthor{A. Hubin et al.}

\address[addr1]{Department of mathematics, University of Oslo,
    \printead{e1} % print email address of "e1"
}
\address[addr2]{Department of mathematics, University of Oslo,
    \printead{e2} % print email address of "e1"
}

\address[addr3]{Department of Medical Statistics (CEMSIIS), Medical University of Vienna,
    \printead{e3}
}
\address[addr4]{Norwegian Computing Center,
    \printead{e4} % print email address of "e1"
}

\end{aug}

\begin{abstract}
Logic regression was developed more than a decade ago as a tool to construct predictors from Boolean combinations of binary covariates. It has been mainly used to model epistatic effects in genetic association studies, which is very appealing due to the intuitive interpretation of logic expressions to describe the interaction between genetic variations. Nevertheless logic regression has (partly due to computational challenges) remained less well known than other approaches to epistatic association mapping. Here we will adapt an advanced evolutionary algorithm called GMJMCMC (Genetically modified Mode Jumping Markov Chain Monte Carlo) to perform Bayesian model selection in the space of logic regression models. After describing the algorithmic details of GMJMCMC we perform a comprehensive simulation study that illustrates its performance given logic regression terms of various complexity. Specifically GMJMCMC is shown to be able to identify three-way and even four-way interactions with relatively large power, a level of complexity which has not been achieved by previous implementations of logic regression. We apply GMJMCMC to reanalyze QTL mapping data for Recombinant Inbred Lines in \textit{Arabidopsis thaliana} and from a backcross population in \textit{Drosophila} where we identify several interesting epistatic effects. The method is implemented in an R package which is available on github.
\end{abstract}

\end{frontmatter}

\section{Introduction}
\label{intro}
Logic regression (not to be confused with logistic regression) was developed as a general tool to obtain predictive models based on Boolean combinations of binary covariates \citep{Rucz}. Its primary application area is epistatic association mapping  as pioneered by \cite{Rucz2004} and \cite{Koop} although already early on  the method was also used in other areas \citep{Keles04, Janes05}.
Important contributions to the development of logic regression were later made by the group of Katja Ickstadt \citep{Fritsch2, Schw}, which also provided a comparison of different implementations of logic regression \citep{Fritsch1}. \cite{SchwRucz} gave a brief introduction with various applications and potential extensions of logic regression.  
Recently a systematic comparison of the performance of logic regression and  a more classical  regression approach based on Cockerham's coding %\citep{Wang2009}
to detect interactions illustrated the advantages of logic regression to  detect epistasic effects in QTL mapping \citep{Malina2014}.  Given the potential of logic regression to detect interpretable interaction effects in a regression setting it is rather surprising that it has not yet become wider addressed in applications.

Originally logic regression was introduced together with likelihood based model selection, where simulated annealing served as a strategy to obtain one ``best'' model \citep[see][for details]{Rucz}. However, assuming that there is one ``best'' model disregards the problem of model uncertainty. Whilst this approach works  well in  simulation studies, it seems to be quite an unrealistic assumption in real world applications, where there often is no ``true'' model. Hence Bayesian model averaging,  which implicitly takes into account model uncertainty, becomes important. 
Bayesian versions of logic regression combined with model exploration include Monte Carlo logic regression (MCLR) \citep{Koop} and the full Bayesian version of logic regression (FBLR) by~\cite{Fritsch2}.  Both MCLR and FBLR use Markov Chain Monte Carlo (MCMC) algorithms for searching through the space of models and parameters. Inference is then based on a large number of models instead of just one model as in the original version of logic regression. MCLR utilizes a geometric prior on the size of the model (defined through the number of logic terms and their complexity). All models of the same size  get the same prior probability while larger models implicitly are penalized. Regression parameters are marginalized out, significantly simplifying computational complexity. 
In contrast FBLR is performed on a joint space of parameters and models. FBLR uses multivariate normal priors for regression parameters, while model size is furnished with a slightly different prior serving similar purposes as the MCLR prior. In case of a large number of binary covariates these MCMC based methods  might require extremely long Markov chains to guarantee convergence which can make them infeasible in practice. Additionally both of them utilize simple Metropolis-Hastings settings which, together with the fact that the search space is often multimodal, increases the probability that they are stuck in local extrema for a significant amount of time.

In this paper we propose a new approach for Bayesian logic regression including model uncertainty. We introduce a novel prior for the topology of logic regression models which is slightly simpler to compute than the one used by MCLR and which still shows excellent properties in terms of controlling false discoveries. We consider two different priors for regression coefficients: Jeffreys prior and the robust g-priors as a state of the art choice for priors of regression coefficients in variable selection problems. 
For Jeffreys prior computing the marginal likelihoods can be performed with the Laplace approximation as in BIC-like model selection criteria. For the robust g-prior the marginal likelihood is efficiently computed using the integrated Laplace approximation~\citep{li2015mixtures}.  

The main contribution of this paper is the proposed search algorithm, named GMJMCMC, which provides a better search strategy for exploring the model space than previous approaches. 
GMJMCMC combines genetic algorithm ideas with the 
mode jumping Markov Chain Monte Carlo (MJMCMC) algorithm~\citep{tjelmeland2001mode,hubin2016efficient}  in order
to be able to jump between local modes in the model space.
After formally introducing logic regression and describing the GMJMCMC algorithm in detail we will present results from a comprehensive simulation study. The performance of GMJMCMC is compared with MCLR and FBLR in case of logistic models (binary responses) and additionally analyzed for linear models (quantitative responses). Models of different complexities are studied which allows us to illustrate the  potential of GMJMCMC to detect higher order interactions. Finally we apply our logic regression approach to perform QTL mapping using two publicly available data sets. The first study is concerned with the hypocotyledonous stem length  in \emph{Arabidopsis thaliana} using Recombinant Inbred Line (RIL) data \citep{Balasub}, the second one considering  various traits from backcross data of \emph{\hyphenation{Dro-so-phila}Drosophila Simulans} and \emph{Drosophila Mauritana} is presented in the web supplement. 
The method is implemented as an R package which is freely available on GitHub at \url{http://aliaksah.github.io/EMJMCMC2016/}, where one can also find examples of further logic regression applications.

\section{Methods}

\subsection{Logic regression}

The method of logic regression \citep{Rucz} was specifically designed for the situation where covariates are binary and predictors are defined as logic expressions operating on these binary variables.  Logic regression can be applied in the context of the generalized linear model (GLM) as demonstrated in \cite{Malina2014}. It can also be easily expanded to the domain of generalized linear mixed models (GLMM), but to keep our presentation as simple as possible we will focus here on generalized linear regression models.

Consider a response  variable $Y \in \R$, together with  $m$ binary covariates
$X_1,X_2,\dots ,X_m$.  Our primary example will be genetic association studies where, depending on the context, each binary covariate, $X_j,\;$ $j\in\{1,2,\dots,m\},\;$ can have a different interpretation. In QTL mapping with backcross design or recombinant inbred lines $X_j$ simply codes the two possible genetic variants. In case of intercross design or in outbred populations different $X_j$ will be used to code dominant and recessive effects \citep[see for example][]{Malina2014}. We will adapt the usual convention that a value 1 corresponds to logical TRUE and a value 0 to logical FALSE where the immediate interpretation in our examples is that a specific marker is associated with a trait or not. 
Each combination of the binary variables $X_j$ with the logical operators $\wedge $ (AND), $\vee $  (OR) and  ${X}^c$ (NOT $X$), is called a logic expression (for example $L=(X_1 \wedge X_2)\vee {X_3^c}$). Following the nomenclature of \cite{Koop} we will refer to logic expressions as \emph{trees}, whereas the primary variables contained in each tree are called \emph{leaves}. The set of leaves of a tree $L$ will be denoted by $v(L)$, that is for the specified example above we have $v(L) = \{X_1, X_2, X_3\}$. 

We will study logic regression in the context of the generalized linear model (GLM, see \cite{McCullagh-Nelder-1989}) of the  form
\begin{eqnarray} \label{eq:LRM}
 Y &\sim& \mathfrak{f}\left(y\mid\mu(\boldsymbol X);\phi\right)\\
 h\left(\mu(\boldsymbol X)\right)&=&\alpha+\sum_{j=1}^{q} \gamma_j\beta_j L_j, \label{eq:LRM2}
 \end{eqnarray} 
where $\mathfrak{f}$ denotes the parametric distribution of $Y$ belonging to the exponential family with mean $\mu(\boldsymbol X)$ and dispersion parameter $\phi$. The function $h$ is an appropriate link function, $\alpha$ and $\beta_j, j \in \{1,...,q\}$ are unknown regression parameters, and $\gamma_j$ is the indicator variable which specifies whether the tree $L_j$ is included in the model. For the sake of simplicity we abbreviate by $\mu(\boldsymbol X)$ the complex dependence of the mean $\mu$ on $\boldsymbol{X}$ via the logic expressions $L_j$ according to \eqref{eq:LRM2}.   Our primary examples are linear regression for quantitative responses and logistic regression for dichotomous responses but the implementation of our approach works for any generalized linear model.

We will restrict ourselves to trees with no more than $C_{max}$ leaves. Consequently the total number of trees $q$ will be finite. The considered models are restricted to include no more than $k_{max}$ trees.
The vector of binary random variables  $M = (\gamma_1,\dots,\gamma_q)$  fully characterizes a model in terms of which logical expressions are included. Here we go along with the usual convention in the context of variable selection that 'model' refers to the set of regressors and does not take into account the specific values of the non-zero regression coefficients. 

\subsubsection{Bayesian model specification}

For a fully Bayesian approach one needs prior specifications for the model topology characterized by the index vector $M$ as well as for the coefficients $\alpha$ and $\beta_j$ belonging to a specific model $M$. This is a common approach in Bayesian model selection, used for example in \citet{clyde2011bayesian} or~\citet{hubin2016efficient}. We start with defining the prior for $M$ by
\begin{align}
p(M)\propto\ &\mathbb{I}\left(|M|\leq k_{max}\right)\prod_{j=1}^q \rho(\gamma_j).\label{eq:modelprior}
\end{align}
Here $|M|=\sum_{j=1}^q\gamma_j$ is the number of logical trees included in the model and $k_{max}$ is the maximum number of trees allowed per model. The factors $\rho(\gamma_j)$ are introduced to give smaller prior probabilities to more complex trees. Specifically we consider 
\begin{align}
\rho(\gamma_j)=\ &  a^{\gamma_jc(L_j)}\label{glmgammaprior}
\end{align}
with $0<a<1$ and $c(L_j)\ge 0$ being a non-decreasing measure for the complexity of the corresponding logical trees.  In case of $\gamma_j =  0$ it holds that  $\rho(\gamma_j) = 1$ and thus the prior probability for model $M$ only consists of the product of $\rho(\gamma_j)$ for all trees included in the model.  
It follows that if $M$ and $M'$ are two vectors only differing in one component, say $\gamma_j'=1$ and $\gamma_j = 0$, then
\[
\frac{p(M')}{p(M)}=a^{c(L_j)}<1
\]  
showing that larger models are penalized more. This result easily generalizes to the comparison of more different models  and provides the basic intuition behind the chosen prior. 

The prior choice implies a distribution for the model size $|M|$ which can be interpreted as a multiple-testing penalty~\citep{Scott}. For $k_{max}=q$ and a constant complexity value on all trees, $|M|$ follows a binomial distribution. With varying complexity measures,  $|M|$ follows the \emph{Poisson binomial} distribution \citep{wang1993number} 
which is a unimodal distribution with
$E[|M|]=\sum_{j=1}^qp_j$ and
$\text{Var}[|M|]=\sum_{j=1}^qp_j(1-p_j)$
where $p_j=a^{c(L_j)}/(1+a^{c(L_j)})$. A truncated version of this distribution is obtained for $k_{max}<q$.

The choices of $a$ and the complexity measure $c(L_j)$ are crucial for the quality of the model prior. 
Let $N(s)$ be the total number of trees having $s$ leaves. 
Choosing
 $a = e^{-1}$ and $c(L_j) = \log N(s_j)$ 
 as long as the number of leaves is not larger than $C_{max}$
 results for $\gamma_j = 1$ in
$$
  a^{c(L_j)} = 
  \frac{1}{N(s_j)}  \; , \quad  s_j\leq C_{max} \;.
$$
Therefore the multiplicative contribution of a specific tree of size $s$ to the model prior will be indirectly proportional to the total number of trees $N(s)$ having $s$ leaves as long as $s \leq C_{max}$. 
Given that $N(s)$ is rapidly growing with the tree size $s$ this choice gives smaller prior probabilities for larger trees. The resulting penalty closely resembles the Bonferroni correction in multiple testing as discussed for example by \cite{BGT08} in the context of modifications of the BIC.

The number $N(s)$ will in practice be difficult to compute. 
To compute a rough approximation of $N(s)$ we ignore logic expressions including the same variable multiple times. Then there are $\binom{m}{s}$ possibilities to select variables. Each variable can undergo logic negation giving $s$ binary choices and furthermore there are $s-1$ logic symbols $(\vee, \wedge)$ to be chosen resulting in $2^{2s-1}$ different expressions. However, due to De Morgan's law half of the expressions provide identical logic regression models. This gives
\begin{equation} \label{NumberOfTrees}
N(s)\approx\binom{m}{s}\ 2^{2s-2}.
\end{equation}
Using this approximation, 
for a model of size $k = |M|$ the full model prior is of the form
\begin{equation} \label{ModelPrior}
P(M) \propto   \mathbb{I}\left(k\leq k_{max}\right)\prod_{r=1}^{k}\frac{\mathbb{I}\left(s_{j_r}\leq C_{max}\right)}{\binom{m}{s_{j_r}}2^{2s_{j_r}-2}} \;,
\end{equation}
where $j_1,\dots,j_{k}$ refer to the $k$ trees of model $M$.

We will next discuss priors for the parameters given a specific model $M$.
 The GLM formulation  \eqref{eq:LRM} includes a dispersion parameter $\phi$, which for example in case of the linear model  is connected with the variance term $\sigma^2$ for the underlying normal distribution. If a GLM has a dispersion parameter then for the sake of simplicity we will adapt the commonly used improper prior \citep{li2015mixtures,bayarri2012criteria}
\begin{align} \label{variance_prior}
\pi(\phi) =& \phi^{-1} \; .
\end{align}
If a GLM does not include a dispersion parameter (like logistic regression) then one simply sets $\phi = 1$.

Concerning the intercept $\alpha$ and the regression coefficients $\beta_j$, where $j \in \{j_1,...,j_{|M|}\}$ correspond to the non-zero coefficients of model $M$,  we will consider two different types of priors, simple Jeffreys priors and robust g-priors.
Jeffreys prior~\citep{jeffreys1946invariant,jeffreys1961theory,gelman2013bayesian} assumes for the parameters of the model an improper prior distribution of the form
\begin{align} \label{Jeffrey}
\pi_{\alpha}(\alpha)\pi_{\beta}(\boldsymbol{\beta})=&|\mathcal{J}_n(\alpha,\boldsymbol{\beta})|^{\frac{1}{2}}\;,
\end{align}
where $\mathcal{J}_n(\alpha,\boldsymbol{\beta})$ is the observed information.

To obtain model posterior probabilities 
one needs to evaluate the marginal likelihood of the model $P(Y\mid M)$ by integrating over all parameters of the model which is often a fairly difficult task. The greatest advantage of Jeffreys prior is that this integral can be approximated  simple and accurate through the   
 Laplace approximation. In case of the Gaussian model choosing Jeffreys prior \eqref{Jeffrey} for the coefficients and the simple prior \eqref{variance_prior} for the variance term yields that the Laplace approximation becomes exact \citep{raftery1997bayesian} and gives a marginal likelihood of the simple form
\begin{align} \label{MargLik_Jeffrey}
P(Y\mid M) \propto & P(Y\mid M,\hat\theta)\ n^{\frac{|M|}{2}} \;,
\end{align}
 where $\hat\theta$ refers to the maximum likelihood estimates of all parameters involved.  On the log scale this exactly corresponds to the BIC model selection criterion \citep{Schwarz} when using a uniform model prior. In case of logistic regression the marginal likelihood under Jeffreys prior becomes approximately \eqref{MargLik_Jeffrey} with an error of order $O(n^{-1})$ \citep{KadaneErr,claeskens_hjort_2008}. \citet{Barber2016} also describe that Laplace approximations of the marginal likelihood yield very accurate results and can be trusted in Bayesian model selection problems.
 
Although there are many situations in which selection based on BIC like criteria works well, within the Bayesian literature using Jeffreys prior for model selection has been widely criticized for not being consistent once the true model coincides with the null model \citep[all $\gamma_j=0$,][]{bayarri2012criteria}. A large number of alternative priors have been studied, see for example \citet{li2015mixtures} who give a comprehensive review on the state of the art of g-priors. In a recent paper \citet{bayarri2012criteria} gave theoretical arguments in case of the linear model recommending the \emph{robust} g-prior, which is consistent in all situations and yields errors diminishing significantly faster than other prior choices. Thus we will introduce the robust g-prior as an alternative to Jeffreys prior.

Our description of robust g-priors  follows  \cite{li2015mixtures} who consider an improper constant prior for the intercept, $P(\alpha) \propto 1$, and a mixture g-prior  for the regression coefficients $\beta_j, j \in \{j_1,...,j_{|M|}\}$ of the form
\begin{align}
P(\boldsymbol \beta\mid g) &\sim  N_{|M|}\left(\boldsymbol 0,g\cdot\phi\mathcal{J}_n({\boldsymbol\beta})^{-1}\right) \;.
\end{align}
Here $\mathcal{J}_n({\boldsymbol\beta})$ is the subblock of the full observed information matrix $\mathcal{J}_n({\alpha,\boldsymbol\beta})$ related to $\boldsymbol{\beta}$ and $g$ itself is assumed to be distributed according to the so called truncated Compound Confluence Hypergeometric (tCCH) prior  
\begin{align}
P\left(\frac{1}{1+g}\right) &\sim tCCH\left(\frac{a}{2},\frac{b}{2},r,\frac{s}{2},v,\kappa\right).
\end{align}

This family of mixtures of g-priors includes a large number of priors discussed in the literature, see  \cite{li2015mixtures} for more details. 
The recommended robust g-prior is a particular case with the following choice of parameters: $$a=1, b=2, r = 1.5,  s=0, v = \frac{n+1}{|M|+1}, \kappa =1 \;.$$ Under this prior specification precise integrated Laplace approximations of the marginal likelihood for GLM are given by~\cite{li2015mixtures}, whilst exact values are available for Gaussian models \citep{li2015mixtures,bayarri2012criteria}.

\subsection{Computing posterior probabilities}

Given prior probabilities for any logic regression  model $M$ the model posterior probability   can be computed according to Bayes formula as
\begin{equation}\label{post}
P(M\mid Y)=\frac{P(Y\mid M) P(M)}{\sum_{M' \in \Omega} P(Y\mid M') P(M')}\;\;,
\end{equation}
where $P(Y\mid M)$ denotes the integrated (or marginal) likelihood for model $M$  and $\Omega$ is the set of all models in the model space. The sum in the denominator involves a huge number of terms and it is impossible to compute all of them. Classical MCMC based approaches (like MCLR and FBLR)  overcome this problem by estimating model posteriors with the relative frequency with which a specific model $M$ occurs in the Markov chain. In case of an ultrahigh-dimensional model space (like in case of logic regression) this is computationally extremely challenging and might require chain lengths which are prohibitive for practical applications.

An alternative approach makes use of the fact that most of the summands in the denominator of~\eqref{post} will be so small that they can be neglected. Considering a subset $\Omega^* \subseteq \Omega$ containing the most important models we can therefore approximate~\eqref{post} by
\begin{equation}\label{Approx_post}
P(M\mid Y)\approx  \tilde P(M\mid Y)=\frac{P(Y\mid M) P(M)}{\sum_{M' \in \Omega^*} P(Y\mid M') P(M')}\;.
\end{equation}
To obtain good estimates we have to search in the model space for those models that contribute significantly to the sum in the denominator, that is for those models with large posterior probabilities or equivalently with large values of $P(Y\mid M) P(M)$. In \cite{FLAB12} specific memetic algorithms were developed to perform the model search for linear regression. Here we will rely upon the GMJMCMC algorithm, which is described in the next section. For now we assume that some method for computing the marginal likelihood $P(Y\mid M)$ is available. The details of such computation depend on the prior specifications of the parameters of a particular model and are given for the examples in the experimental sections.

Based on model posterior probabilities one can easily obtain an estimate of the posterior probability for a logic expression $L_j$ to be included in a model (also referred to as the marginal inclusion probability) by
\begin{equation}\label{Approx_post_eta}
\tilde P(L_j\mid Y) = \sum_{M\in \Omega^*: \gamma_j = 1} \tilde P(M\mid Y).\footnote{Here by $P(L_j\mid Y)$ we mean $P(\gamma_j=1\mid Y)$.}
\end{equation}
Inference on trees can then be performed by means of selecting those trees with a posterior probability being larger than some threshold probability $\pi_C$. In case of exploratory studies where the main aim is to discover many potentially interesting features to be explored in further studies it can be reasonable to use low threshold values on  $\tilde P(L_j\mid Y)$. High threshold values can be used if false discoveries need to be avoided. In general the threshold can be specified through a decision theoretic framework, including the aim of controlling  false discovery rates, see~\citep{wakefield2007bayesian}.

A threshold of 0.5 corresponds to  the median probability model of~\citet{barbieri2004optimal} which under certain circumstances has greater predictive power than the most probable model. 
However, one of the criteria for the median probability model to be optimal in the linear Gaussian case, the graphical model structure criterion, will not always be valid in cases were one makes restrictions on the number of trees that can be included. The graphical model structure criterion requires that the median probability model results in a legal model. Consider the case with three covariates $x_1,x_2,x_3$ but with $k_{max}=2$ and the posterior probabilities for models $\bm\gamma=(1,1,0)$, $\bm\gamma=(1,0,1)$ and $\bm\gamma=(0,1,1)$ each equal to $1/3$. Then all marginal inclusion probabilities are  $2/3$ and the median probability model includes all variables which then has a model size larger than $k_{max}$.  The median probability model can however still be a useful model to consider even in cases where the optimality results do not apply.

\subsection{The GMJMCMC algorithm}

To fix ideas consider first a variable selection problem with $q$ potential covariates to enter a model. Recall that $\gamma_j$ needs to be 1 if the $j$-th variable is to be included into the model and 0 otherwise. A model $M$ is thus specified by the vector $\BB\gamma=(\gamma_1,...,\gamma_q)$ and the general model space $\Omega$ is of size $2^q$. If this discrete model space is multimodal in terms of model posterior probabilities then simple MCMC algorithms typically run into problems by staying for too long in the vicinity of local maxima.  Recently, the mode jumping MCMC procedure (MJMCMC) was proposed by~\cite{hubin2016efficient} to overcome this issue in a model selection setting. 

MJMCMC is a proper MCMC algorithm equipped with the possibility to jump between different modes within the discrete model space. 
The key to the success of MJMCMC is the generation of good proposals of models which are not too close to the current state. This is achieved by first making a large jump (changing many model components) and then performing local optimization within the discrete model space to obtain a proposal model. Within a Metropolis-Hastings setting a valid acceptance probability is then constructed using symmetric backward kernels, which guarantees that the resulting Markov chain is ergodic and has the desired limiting distribution \citep{tjelmeland2001mode,hubin2016efficient}. 

The MJMCMC algorithm requires that all of the covariates defining the model space are known in advance and are all considered at each iteration of the algorithm. In case of logic regression the covariates are trees and a major problem in this setting is that it is quite difficult to fully specify the space $\Omega$. In fact it is even difficult to specify $q$, the total number of feasible trees. To solve this problem we present an adaptive algorithm called  Genetically Modified MJMCMC (GMJMCMC), where
MJMCMC is embedded in the iterative setting of a genetic algorithm. In each iteration only a given set $\mathcal{S}$ of trees (of fixed size $d$)  is considered. Each $\mathcal{S}$ then induces a separate \textit{search space} for MJMCMC. In the language of genetic algorithms $\mathcal{S}$ is the \textit{population}, which dynamically evolves to allow MJMCMC exploring different reasonable parts of the unfeasibly large total search space. %The resulting algorithm is similar to feature engineering~\citep{FIPAP} and allows to consider combinations of covariates that can be adapted throughout the search.
   
To be more specific, we consider different populations  $\mathcal{S}_1,\mathcal{S}_2,...$ where each $\mathcal{S}_t$ is a set of $d$ trees.  
For each given population a fixed number of MJMCMC steps is performed. Since the
MJMCMC algorithm is specified in full detail in~\cite{hubin2016efficient}, we will concentrate here on describing the evolutionary dynamics yielding subsequent populations $\mathcal{S}_t$. 
Utilization of the approximation~\eqref{Approx_post} in combination with exact or approximated marginal likelihoods allows us to compute posterior probabilities for all models in $\Omega^*$ which have been visited at least once by the algorithm. Consequently we do not need a
proper MCMC (an algorithm with convergence towards the target distribution) which is needed if model posterior probabilities are estimated by the relative frequency of how often a model has been visited. 
In principle it is possible to construct a proper MCMC algorithm which aims at simulating from extended models of the form $P(M,\mathcal{S}\mid Y)$ having $P(M\mid Y)$ as a stationary distribution. This version of the algorithm is considered in~\citep{hubin2018deep} where the main idea is to  perform both forward and backward swaps between populations in order to obtain a reversible Markov chain. 

The algorithm is initialized by first running MJMCMC for a given number of iterations $N_{init}$ on the set of all binary covariates $X_1,...,X_m$ as potential regressors, but not including any interactions.   The first $d_1<d$ members of population $\mathcal{S}_1$ are then defined to be the $d_1$ covariates with largest marginal inclusion probability. In our current implementation we select the $d_1$ leaves which have marginal posterior probabilities (estimated from the first $N_{init}$ iterations) larger than $\rho_{min}$, thus $d_1$ is not pre-specified but is obtained in a data driven way. 
 For later reference we denote this set of $d_1$ leaves by $\mathcal{S}_0$. 
The remaining $d-d_1$ members of  $\mathcal{S}_1$ are obtained by forming logic expressions from the leaves of $\mathcal{S}_0$ where trees are generated randomly by means of the crossover operation described below. In practice one first has to choose some $k_{max}$ which will depend on the expected number of trees to enter the model in the problem one studies. The choice of $d$ can then be guided by the results of Theorem \ref{th:GMJMCMC} given below. 

After $\mathcal{S}_1$ has been initialized MJMCMC is performed for a fixed number of iterations  $N_{expl}$ before the next population $\mathcal{S}_2$ is generated. This process is iterated for $T_{max}$ populations $S_t,  t \in \{1,...,T_{max}\}$. 
The $d_1$ input trees from the initialization procedure remain in all populations $\mathcal{S}_t$ throughout our search. Other trees from the population $\mathcal{S}_t$ with low marginal inclusion probabilities (below a threshold $\rho_{min}$) will be substituted by trees which are generated by crossover, mutation and reduction operators to be described in more detail below.

Let $D_t$ be the set of trees to be deleted from $\mathcal{S}_t$.
Then $|D_t|$ replacement trees must be generated instead. Each replacement tree is generated randomly by a \textit{crossover} operator with probability $P_c$ and by a \textit{mutation} operator with probability $P_m=1-P_c$. A \textit{reduction} operator is applied if \textit{mutation} or \textit{crossover} gives a tree larger than the maximal tree size $C_{max}$.
 
\noindent{\textbf{Crossover:}} 
Two \emph{parent trees} 
are selected from $\mathcal{S}_t$ with probabilities proportional to the approximated \hyphenation{mar-gi-nal}marginal inclusion probabilities of trees in $\mathcal{S}_t$. Then each one of the parents is inverted with probability $P_{not}$ by the logical not $^c$ operator, before they are combined with a $\wedge$ operator with probability $P_{and}$ and with a $\vee$ operator otherwise. 
Hence the \hyphenation{cross-over}crossover operator gives trees of the form $L_{j_1}\wedge L_{j_2}$ or  $L_{j_1}\vee L_{j_2}$ where
either $L_{j_i}$ or $L_{j_i}^c$ is in $\mathcal{S}_t$ for $i=1,2$.
 
\noindent{\textbf{Mutation:}} 
One parent tree is selected from $\mathcal{S}_t$ with probability proportional to the approximated marginal inclusion probabilities of trees in $\mathcal{S}_t$, whilst the other parent tree is selected uniformly from  the set of $m - d_1$ leaves which did not make it into the initial population $\mathcal{S}_0$. Then just like for the crossover operator each of the parents is inverted with probability $P_{not}$ by the logical not $^c$ operator, before they are combined  with a $\wedge$ operator with probability $P_{and}$ and with a $\vee$ operator otherwise. The mutation operator gives trees of the form $L_{j_1}\wedge X$ or  $L_{j_1}\vee X$ where
either $L_{j_1}$ or $L_{j_1}^c$ is in $\mathcal{S}_t$  and $X$ or $X^c$ is in $D_0$.
 
\noindent{\textbf{Reduction:}}  A new tree is generated from a tree by deleting a subset of leaves, where each leave has a probability of $\rho_{del}$ to be deleted. The pruning of the tree is performed in a natural way meaning that the 'closest' logical operators of the deleted leaves are also deleted. If the deleted leave is not on the boundaries of the original tree the operation is resulting in obtaining two separated subtrees. The resulting subtrees are then combined in a tree with a $\wedge$ operator with probability $P_{and}$ or with a $\vee$ operator otherwise. 
 
For all three operators it holds that if the newly generated tree is already present in $\mathcal{S}_t$ then it is not considered for  $\mathcal{S}_{t+1}$ but rather a new replacement tree is proposed instead. 
The pseudo-code \textbf{Algorithm 1} describes the full GMJMCMC algorithm. For each iteration $t$ the initial model for the next MJMCMC run is constructed by randomly selecting trees from $\mathcal{S}_t$ with probability $P_{init}$. 
For the final population $\mathcal{S}_{T_{max}}$,  MJMCMC is run until $M_{fin}$ unique models are visited (within $\mathcal{S}_{T_{max}}$). $M_{fin}$ should be sufficiently large to obtain good MJMCMC based approximations of the posterior parameters of interest based on the final search space $\mathcal{S}_{T_{max}}$.

\begin{algorithm}[h]
\caption{GMJMCMC}\label{gMJMCMCalg}
\begin{algorithmic}[1]
\item Run the MJMCMC algorithm for $N_{init}$ iterations on $X_1,...,X_m$ and define  $\mathcal{S}_0$ as the set of $d_1$  variables among them with the largest estimated marginal inclusion probabilities. 
\State Generate $d-d_1$ trees by randomly selecting crossover operations of elements from $\mathcal{S}_0$ and add those trees to the set $\mathcal{S}_0$ to obtain $\mathcal{S}_1$. 
\State Run the MJMCMC algorithm within search space $\mathcal{S}_1$.
\For{$t=2,...,T_{max}$} 
\State Delete trees within $\mathcal{S}_{t-1}\backslash\mathcal{S}_0$ which have estimated inclusion probabilities less than $\rho_{min}$.
\State Add new trees which are generated by crossover, mutation or reduction operators until the having again a set of size $d$, which becomes  $\mathcal{S}_t$.
\State Run the MJMCMC algorithm within search space $\mathcal{S}_t$.
\EndFor
\end{algorithmic}
\end{algorithm}

The following result is concerned with consistency of probability estimates of GMJMCMC when the number of iterations increases.
\begin{thm}\label{th:GMJMCMC}
Assume $\Omega^*$ is the set of models visited through the GMJMCMC algorithm where
 $d-d_1\ge k_{max}$. Assume further the marginal likelihoods are calculated without errors. Then the model estimates based on~\eqref{Approx_post} will converge to the true model  probabilities as the number of iterations $T_{\max}$ goes to $\infty$.
\end{thm}
\begin{proof}
Note that the approximation~\eqref{Approx_post} will provide the exact answer if $\Omega^*=\Omega$. It is therefore enough to show that the algorithm in the limit will have visited all possible models. Since $\mathcal{S}_0$ is generated in the first step and never changed, we will consider it to be fixed.

Define $M_{S_t}$ to be the last model visited by the MJMCMC algorithm on search space $\mathcal{S}_t$. 
Then the construction of $\mathcal{S}_{t+1}$ only depends on 
$(\mathcal{S}_t,M_{S_t},\boldsymbol X)$ while $M_{\mathcal{S}_{t+1}}$ only depends on $\mathcal{S}_{t+1}$. Therefore $\{(\mathcal{S}_t,M_{\mathcal{S}_t},\boldsymbol X)\}$ is a Markov chain. Assume now $\mathcal{S}$ and  $\mathcal{S}'$ are two populations differing in one component with $L\in\mathcal{S}$, $L'\in\mathcal{S}'$, $L\neq L'$.  Define $L_{sub}$ to be any tree that is a subtree of both $L$ and $L'$ (where a subtree is defined as a tree which can be obtained by reduction)  and $\mathcal{S}_{sub}$ to be the search space where $L$ is substituted with $L_{sub}$
in $\mathcal{S}$. Then it is possible to move from $S$ to $S_{sub}$ in $l$ steps using first \emph{mutations} and \emph{crossovers} to grow a tree $L^*$ of size larger than $C_{max}$, which can undergo \emph{reduction} (note that although only trees that have low enough estimated marginal inclusion probabilities can be deleted, there will always be a positive probability that marginal inclusion probabilities are estimated to be smaller than the threshold $\rho_{min}$) to get to $L_{sub}$. Further, assuming the difference in size between $L_{sub}$ and $L'$ is $r$, a move from $S_{sub}$ to $S'$ can be performed by $r$ steps of \emph{mutations} or \emph{crossovers}. Two search spaces which differ in $s$ trees can be reached by $s$ combinations of the moves described above.
Since also any model within a search space can be visited, the Markov chain $\{(\mathcal{S}_t,M_{\mathcal{S}_t},\boldsymbol X)\}$ is irreducible. Since the state space for this Markov chain is finite, it is also recurrent, and there exists a stationary distribution with positive probabilities on every model. Thereby, all states, including all possible models of maximum size  $d$, will eventually be visited.

When  $d_1>0$, some restrictions on the possible search spaces are introduced. However, when $d-d_1\ge k_{max}$, any model of maximum size $k_{max}$ \emph{will} eventually be visited. 
\end{proof}
\paragraph{Remark 1} If  $d-d_1 < k_{max}$, then every model of size up to $d-d_1$  plus some of the larger models will eventually be visited, although the model space will get some additional constraints. In practice it is more important that $d-d_1\ge k^*$, where $k^*$ is the size of the true model. Unfortunately neither $k^*$ nor $d_1$ are known in advance, and one has to make reasonable choices of $k_{max}$ and $d$ depending on the problem one analyses.\hfill\qedsymbol
\paragraph{Remark 2} The result of Theorem \ref{th:GMJMCMC} relies on exact calculation of the marginal likelihood $P(Y\mid M)$.
Apart from the linear model, the calculation of $P(Y\mid M)$ is typically based on an approximation, giving similar approximations to the model probabilities. How precise these approximations are will depend on the type of method used. The current implementation includes Laplace approximations, integrated Laplace approximations, and integrated nested Laplace approximations. In principle other methods based on MCMC outputs \citep{chib1995marginal,chib2001marginal} could  be incorporated relatively easily
 resulting however in longer runtimes. 

\subsubsection*{Parallelization}
Due to our interest in exploring as many \emph{unique}  high quality models as possible and doing it as fast as possible, running multiple parallel chains is likely to be computationally beneficial compared to running one long chain. The process can be embarrassingly parallelized into $B$ chains using several CPUs, GPUs or clusters. If one is mainly interested in model probabilities, then Equation~\eqref{Approx_post} can be directly applied with $\Omega^*$ now being the set of unique models visited within all runs. However, we suggest a more memory efficient approach. If some statistic $\Delta$
is of interest, one can utilize the following posterior estimates based on weighted sums over individual runs:
\begin{equation}\label{weighted_sum}
\tilde{P}(\Delta\mid Y) = \sum_{b=1}^Bw_b\tilde{P}_b(\Delta\mid Y)\;.
\end{equation}
Here $w_b$ is a set of weights which will be specified below and $\tilde{P}_b(\Delta\mid Y)$ are the posteriors obtained with formula~\eqref{Approx_post_delta} from run $b$ of GMJMCMC. 

Due to the irreducibility of the GMJMCMC procedure it holds that for $\sum_bw_b=1$ we obtain $\lim_{T_{max}\rightarrow \infty}\tilde{P}(\Delta\mid Y) = P(\Delta\mid Y)$ where $T_{max}$ is the number of iterations within each run. Thus for any set of normalized weights the approximation $\tilde{P}(\Delta\mid Y)$ converges to the true posterior probability ${P}(\Delta\mid Y)$. Therefore in principle any normalized set of weights $w_b$ would work, like for example $w_b = \frac{1}{B}$. However, uniform weights have the disadvantage to potentially give too much weight to posterior estimates from  chains that have not quite converged. In the following heuristic improvement $w_b$ is chosen to be proportional to the posterior mass detected by run $b$, 
\begin{align*}
w_b=&\frac{\sum_{M' \in \Omega^*_b} P(Y\mid M') P(M')}{\sum_{b=1}^B\sum_{M' \in \Omega^*_b} P(Y\mid M') P(M')}\; .
\end{align*}
This choice  indirectly penalizes chains that cover smaller portions of the model space. When estimating posterior probabilities using these weights we only need, for each run,  to store the following quantities: $\tilde{P}_b(\Delta\mid Y)$ for all statistics $\Delta$ of interest and $s_b = \sum_{M' \in \Omega^*_b} P(Y\mid M') P(M')$ as a \textit{'sufficient'} statistic of the run. There is no further need of data transfer between processes. 

Alternatively (as mentioned above) one might use~\eqref{Approx_post_delta} directly to approximate $P(\Delta\mid Y)$ based on the totality $\Omega^*$ of unique models explored through all of the parallel chains. This procedure might give in some cases slightly better precision  than the weighted sum approach~\eqref{weighted_sum}, but it is still only asymptotically unbiased. Moreover keeping track of all models visited by all chains requires significantly more storage in the quick memory and RAM and requires significantly more data transfers across the processes. 
Consequently this approach is not part of the current implementation of GMJMCMC.

The consistency result of Theorem 1 also holds in case of the suggested embarrassing parallelization. Moreover it holds that even when the number of iterations per chain is finite that letting the numbers of chains $B$ go to infinity yields consistency of the posterior estimates as shown in  Theorem A.1 in the web supplement. The main practical consequence is that running more chains in parallel allows for having a smaller number of iterations within each thread.

\paragraph{Choice of algorithmic parameters} Apart from the number of parallel chains, the GMJMCMC algorithm relies upon the choice of a number of tuning parameters which were described above. Section A of the web supplement presents the values that were used in the following simulation study and in real data analysis.

\section{Experiments}

\subsection{Simulation study}

The GMJMCMC algorithm was evaluated in a simulation study divided into two parts. The first part considered three scenarios (numbered 1-3) with binary responses and the second part three scenarios (4-6) with  quantitative responses. For each scenario we generated $N = 100$ datasets according to a regression model described by Equations~\eqref{eq:LRM} and ~\eqref{eq:LRM2}  with $n=1000$ observations and $p=50$ binary covariates. The covariates were assumed to be independent and were simulated for each simulation run  as $X_{j}\sim \text{Bernoulli}(0.3)$ for $j \in \{1,\dots,50\}$ in the first two scenarios and as $X_{j}\sim \text{Bernoulli}(0.5)$ for $j \in \{1,\dots,50\}$ in the last four scenarios. All computations were performed on the Abel cluster\!
\footnote{The Abel cluster node (\url{http://www.uio.no/english/services/it/research/hpc/abel/}) with 16 dual Intel E5-2670 (Sandy Bridge, 2.6 GHz.) CPUs and 64 GB RAM under 64 bit CentOS-6  is a shared resource for research computing.}.       

For Scenarios 3, 5 and 6 the effect sizes ($\beta_j$'s) for higher order interactions might seem unrealistically large compared to real applications. To obtain more realistic scenarios with moderate effect sizes and still sufficient power to detect larger trees one would have  to increase the sample sizes. However, this would be quite challenging computationally for a simulation study. In the section on sensitivity analysis additional simulations for Scenario 5  illustrate  which effect sizes are needed with a sample size of $n=1000$ for GMJMCMC to detect trees of different size. Furthermore,  we demonstrate  that increasing the sample size by a factor  $10$ and reducing the effect sizes by a factor $1 / \sqrt{10}$ yields approximately the same power. This relationship indicates which sample sizes would be necessary in practice to detect higher order interactions with smaller effect sizes.

\subsubsection*{Binary responses} The responses of the first three scenarios were sampled as modes of Bernoulli random variables with individual success probability $\pi$   specified according to
\begin{align*}
{\bf S. 1:\ } 
\text{logit}(\pi) =&  -0.7 + L_1 +  L_2 +  L_3\\
{\bf S. 2:\ } 
\text{logit}(\pi) = &-0.45+0.6\ L_1 + 0.6\ L_2+ 0.6\ L_3\\
{\bf S. 3:\ } 
\text{logit}(\pi)   = & \ \ \ \ \ 0.4 - 5\ L_1+ 9\ L_2-9\ L_3
\end{align*}
where the corresponding logic expressions are provided in Table \ref{Tab:LogReg}.
The first two scenarios with models including only two-way interactions were copied from \cite{Fritsch2} except that we deliberately did not specify the trees  in lexicographical order. The reason for this is that for some procedures (like stepwise search) it might be an algorithmic advantage if the effects are specified in a particular order. The second scenario is slightly more challenging than the first one due to the smaller effect sizes. The third scenario is  more demanding with a model including three-way and four-way interactions. As mentioned above the corresponding regression coefficients were chosen rather large to make sure that these higher order trees can be detected for the given sample size. In practice when interested in smaller effects one would need larger sample sizes.

For the binary response  scenarios GMJMCMC was compared with FBLR~\citep{Fritsch2} and MCLR~\citep{Koop}, where GMJMCMC was run with Jeffreys prior as well as with the robust g-prior. 
For GMJMCMC the default setting of the maximal number of leaves per tree is $C_{max}=5$. For Scenarios 1 and 2 we additionally report the results for $C_{max} = 2$, which were the values used in the original study of \cite{Fritsch2} and which we also used here for MCLR and FBLR.  For Scenario 3 we set $C_{max}=5$ for all three approaches. The maximal number of trees per model was set to $k_{max} = 10$ for GMJMCMC and FBLR whereas for MCLR it is only possible to specify a maximum of $k_{max} = 5$. This is apparently due to the complexity of  prior computations in MCLR.  Apart from the specification of $C_{max}$ and $k_{max}$  we used for all 3 algorithms their default priors.  In all scenarios we used $d = 15$ for the population size in GMJMCMC.

 GMJMCMC was run until up to $1.6\times 10^6$ models were visited in the first two scenarios and up to $2.7 \times 10^6$ models were visited for the third scenario (divided approximately equally on 32 parallel runs). The length of the Markov chains for FBLR and MCLR were chosen to be $2\times 10^6$ for the first two scenarios and $3 \times 10^6$ for the third scenario. 

By default a tree is classified as detected if the (estimated) marginal inclusion probability is larger than 0.5. This corresponds to the median probability model of~\citet{barbieri2004optimal}.
To evaluate the performance of the different algorithms we estimated the following metrics:
\begin{description}
\item [{\textnormal{\textit{Individual power}}}] - the power to detect a particular tree from the data generating model; \item [\textnormal{\textit{Overall power}}]- the average power over all true trees; 
\item [\textnormal{\textit{FP}}] - the expected number of false positive trees; 
\item [\textnormal{\textit{FDR}}] - the false discovery rate of trees; 
\item [\textnormal{\textit{WL}}] - the total number of wrongly detected leaves. 
\end{description}
Further computational details are given in Section B.1 of the web supplement.

\begin{table}
\centering

\begin{tabular}{lrrcc}
\hline \noalign{\smallskip}
  &FBLR&MCLR&\multicolumn{2}{c}{GMJMCMC}\\
\noalign{\smallskip}  \hline \noalign{\smallskip}
{\bf Scenario 1}&\multicolumn{2}{c}{}&\textbf{Jef.}&\textbf{R. g}\\ 
$L_1 = X_1^c\wedge X_{4}$& 0.30&$\leq$ 0.67& 0.99 (0.97) & 1.00 (0.98)\\
$L_2 = X_5\wedge X_9$&0.42&$\leq$ 0.61    & 0.99 (1.00)& 0.96 (0.95)\\
$L_3 = X_{11}\wedge X_8$&0.33&$\leq$ 0.59 & 0.95 (0.91)&0.53 (0.77)\\
Overall Power&0.35&$\leq$ 0.62      & 0.98 (0.96)&0.84 (0.90)\\
FP&3.88&$\geq 2.70$            & 0.08 (0.25)& 1.01 (0.63)\\
FDR&0.77&$\geq 0.06$         & 0.03 (0.06)&0.25 (0.16)\\
WL&1&0&0 (0)&0 (0)\\
\hline \noalign{\smallskip}
{\bf Scenario 2}&\multicolumn{3}{c}{}\\ 
$L_1 = X_1^c\wedge X_{4}$&0.32&$\leq$ 0.66 & 0.98 (0.97)&0.98 (0.97)\\
$L_2 = X_5\wedge X_9$&0.40&$\leq$ 0.67     & 0.99 (0.99)&0.94 (0.96)\\
$L_3 = X_{11}\wedge X_8$&0.37&$\leq$ 0.60  & 0.96 (0.86)&0.54 (0.76)\\
Overall Power&0.36 &$\leq$ 0.64      & 0.98 (0.94)&0.82 (0.90)\\
FP&3.83 &$\geq 2.58$            & 0.10 (0.38)& 1.08 (0.66)\\
FDR&0.75&$\geq 0.06$          & 0.03 (0.09&0.27 (0.16)\\
WL&1&1&0 (0)&0 (0)\\
\hline \noalign{\smallskip}
{\bf Scenario 3}&\multicolumn{3}{c}{}\\ 
$L_1 = X_2\wedge X_9$&0.93&$\leq 0.93$ & 1.00 & 1.00\\
$L_2 = X_{7}\wedge X_{12}\wedge X_{20}$&0.04&$\leq 0.67$& 0.91& 0.56\\
$L_3 = X_{4}\!\wedge\! X_{10}\!\wedge\! X_{17}\!\wedge\! X_{30}\!\!$&0.00&$\leq 0.19$& 1.00& 0.56\\
Overall Power&0.32&$\leq 0.60$& 0.97& 0.71\\
FP&6.40&$\geq 2.98$ & 0.15 & 1.74\\
FDR&0.54&$\geq 0.06$& 0.04 & 0.39\\
WL&90&72& 1& 0\\
\hline\\
\end{tabular}
\caption{\label{Tab:LogReg}Results for the three simulation scenarios for binary responses. Power for individual trees, overall power,  expected number of false positives (FP) and FDR are compared between  FBLR, MCLR and  GMJMCMC using either Jeffreys prior (Jef.) or the robust g-prior (R.g.). For GMJMCMC the default $C_{max} = 5$ is used. For the first two scenarios we also present results for $C_{max}=2$ (inside parentheses) corresponding to the parameters used by MCLR and FBLR. All algorithms were tuned to use approximately the same computational resources. In case of MCLR we can only provide upper bounds for the power and lower bounds for FP. We also report the total number of wrongly detected leaves (WL) over all simulation runs.} 
\end{table}

A  summary of the results for the first three simulation scenarios is provided in Table~\ref{Tab:LogReg}. In all three scenarios, MCLR performed better than FBLR, even when taking into account the positively biased summary statistics of MCLR (see Section B.1 in the web supplement). On the other hand,  GMJMCMC clearly outperformed MCLR and FBLR both in terms of power and in terms of controlling the number of false positives, where using Jeffreys prior gave slightly better results than using the robust g-prior.

In the first two scenarios GMJMCMC with Jeffreys prior worked almost perfectly both for $C_{max} = 5$ and $C_{max} = 2$. In the few instances where it did not detect the true tree it reported instead the two corresponding main effects. Note however that in case of $C_{max} = 5$ there were several instances where GMJMCMC detected $L_i^c \wedge L_j^c$ with $(1 \leq i < j \leq 3)$, which according to De Morgan's law is equivalent to $L_i + L_j$ and was therefore counted as true positive both for $L_i$ and $L_j$. GMJMCMC with the robust g-prior had a few more instances where pairs of singletons were reported instead of the correct two-way interaction, especially when $C_{max} = 5$ was used. FBLR and MCLR were also good at detecting the true leaves in these  simple scenarios, but GMJMCMC was much better in terms of identifying the exact logical expressions.

The third scenario is more complex than the previous ones but nevertheless GMJMCMC with Jeffreys prior performed almost perfectly.  GMJMCMC with the robust g-prior had more difficulties to correctly identify the three-way and four-way interaction.
Both FBLR and MCLR had  severe problems to detect the true logic expressions and they also reported a considerable number of wrongly detected leaves. For a more in depth discussion of these simulation results we refer to Section B.1 of the web supplement.

\subsubsection*{Continuous responses} Responses were simulated according to a Gaussian distribution with error variance $\s^2 = 1$ and the following three models for the expectation:  
\begin{align*}
{\bf S.4:\ }  E(Y) = 1
+& 1.43\ L_1
+ 0.89\ L_2
+ 0.7\ L_3 \\
{\bf S.5:\ }  E(Y) = 1
+& 1.5\ L_1 
+ 3.5\ L_2 
+ 9\ L_3
+ 7\ L_4  \\
{\bf S.6:\ } E(Y) = 1
+& 1.5\ L_1
+ 1.5\ L_2
+ 6.6\ L_3
+ 3.5\ L_4\\
 +& 9\ L_5  + 7\ L_6
+ 7\ L_7
+ 7\ L_8 
\end{align*}
The logic expressions used in the three different scenarios are provided in Table \ref{Tab:LinReg}.  
Scenario 4 is similar to the first two scenarios for binary responses and contains only two-way interactions. The models of the last two scenarios both include trees of size 1 to 4, where Scenario 5 has one tree of each size. Scenario 6 is the most complex one with two trees of each size, resulting in a model with 20 leaves in total.

\begin{table}[t!]
\centering
\begin{tabular}{lcc}
%\hline
% &GMJMCMC & &GMJMCMC\\ 
\hline \noalign{\smallskip}
{\bf Scenario 4}& \textbf{Jeffreys} & \textbf{Robust g}     \\ 
$L_1 = X_5\wedge X_9$&1.00 &1.00     \\
$L_2 = X_8 \wedge X_{11}$&0.99 &1.00\\
$L_3 = X_1 \wedge X_4$&0.97 &0.98  \\
Overall Power& 0.99  &0.99   \\
FP&0.01 &0.00               \\
FDR&0.005 &0.00            \\
WL&0   &0            \\
\hline \noalign{\smallskip}
{\bf Scenario 5}& \textbf{Jeffreys} & \textbf{Robust g} \\ 
$L_1 = X_{37}$&1.00 &1.00         \\
$L_2 = X_2\wedge X_9$&1.00&0.99 \\
$L_3 = X_7 \wedge X_{12}\wedge X_{20}$ &0.96&1.00 \\
$L_4 = X_4 \wedge X_{10}\wedge X_{17}\wedge X_{30} $&0.89&0.90\\
Overall Power& 0.96 & 0.97\\
FP &0.37&0.28\\
FDR&0.06&0.04 \\
WL&2&5 \\
\hline \noalign{\smallskip}
{\bf Scenario 6}& \textbf{Jeffreys} & \textbf{Robust g} \\ 
$L_1 = X_7$& 0.95 & 0.99 \\
$L_2 = X_8$&0.98 & 0.99 \\
$L_3 = X_2 \wedge X_9$& 0.98& 0.99\\
$L_4 = X_{18}\wedge X_{21}$&0.96&0.95\\
$L_5 = X_1 \wedge X_3 \wedge X_{27}$&1.00 & 1.00\\
$L_6 = X_{12}\wedge X_{20}\wedge X_{37}$&0.95 & 0.96\\
$L_7 = X_4\wedge X_{10}\wedge X_{17}\wedge X_{30}$&0.32 & 0.45\\
$L_8 = X_{11}\wedge X_{13}\vee X_{19}\wedge X_{50}$ &0.21 (0.93) & 0.16 (0.85)\\
Overall Power&0.79 (0.88) & 0.81 (0.90)\\
FP&4.28 (2.05)&4.24 (1.96)\\
FDR&0.38 (0.19)&0.36 (0.16)\\
WL&3&7\\
\hline\\
\end{tabular}
\caption{\label{Tab:LinReg}Results for the three simulation scenarios for linear regression. Power for individual trees, overall power,  expected number of false positives (FP), FDR and the total number of wrongly detected leaves (WL) are given for parallel GMJMCMC. The four estimates in parentheses for Scenario 6 refer to results obtained when counting an equivalent logic expression of $L_8$ as true positive as explained in the text.} 
\end{table}

 For scenarios with Gaussian observations we were only able to study the performance of GMJMCMC since the other approaches cannot handle continuous responses (MCLR has an implementation but that did not work properly).  
For these scenarios the settings of GMJMCMC were adapted to the increasing complexity of the model. We used $k_{max} = 10, 10$ and $20$,  and  $d =15, 20$ and $40$, respectively,  for the three scenarios thus allowing for models larger than  twice the size of the data generating model and populations at least twice the size of the number of correct leaves involved.   Furthermore,  the total number of models visited by GMJMCMC before it stopped was increased  to  $3.5 \times 10^6$ for Scenario 6. $C_{max}$ is set to 5 for all three of these scenarios. Otherwise all parameters of GMJMCMC were set as described for the binary responses.

Table \ref{Tab:LinReg} summarizes the results and further details are provided in Section B.2  of the web supplement. Scenario 4 illustrates that given a sufficiently large sample size GMJMCMC can reliably detect two-way interactions  with  effect sizes smaller than one standard deviation. Both Jeffreys prior and the robust g-prior worked almost perfectly in terms of power. In this simple scenario even the type I error was almost perfectly controlled with false discovery rates equal to 0.005 for Jeffreys prior and 0 for the robust g-prior. Interestingly the only false discovery over all 100 simulation runs was of the form  $X_1 \wedge X_4  \vee X_8 \wedge X_{11}$ and is equal to $L_3 \vee L_2$. One might argue to which extent such a combination of trees should actually be counted as a false positive, a question which is further elaborated in Section B.2 of the web supplement and in the Discussion section.

The remaining two scenarios are  way more complex due to the higher order interaction terms involved.  In Scenario~5  the power to detect any of the four trees was  very large, with only slightly smaller power for the four-way interaction. 
The robust g-prior had only a rather small advantage compared with Jeffreys prior both in terms of power (overall 97\% against 96\%) and in terms of type I error (FDR of 4\% against 6\%). For both priors the majority of false positive results were connected to detecting subtrees of true trees and in all simulation runs there were only 2 wrongly detected leaves for Jeffreys prior and 5 wrongly detected leaves for the robust g-prior.

For the last scenario we again observed large power for all true trees up to order three.  For the final two expressions $L_7$ and $L_8$ of order four the results became slightly more ambiguous with power estimated to 0.32 and 0.21, respectively, for Jeffreys prior and 0.45 and 0.16 for the robust g-prior. However, among the false positive detections we very often found the expressions $X_{11}\wedge X_{13}$, $X_{19}\wedge X_{50}$ as well as $X_{11}\wedge X_{13}\wedge X_{19}\wedge X_{50}$. In fact in 72 simulation runs for Jeffreys prior and 69 simulation runs for the robust g-prior all of these three expressions were detected. According to the logic equivalence
$$
L_8 = X_{11}\wedge X_{13} + X_{19}\wedge X_{50} - X_{11}\wedge X_{13}\wedge X_{19}\wedge X_{50}
$$ 
one might actually consider these findings as true positives.  The numbers in  parentheses in Table~\ref{Tab:LinReg} were  based on taking such similarities into account, resulting in much higher power.  Among the remaining false positive detections more than two thirds were subtrees of true trees or trees with misspecified logical operators but consisting of leaves corresponding to a true tree. Thus again the vast majority of false detections points towards true epistatic effects where the exact logic expression was not identified. 
Interestingly like in Scenario 5 GMJMCMC with the robust g-prior detected again a larger number of wrong leaves than with Jeffreys prior.

\subsubsection*{Sensitivity analysis}

\begin{figure}[h!]
\centering
\begin{minipage}[t]{0.48\linewidth}
1) Regression coefficient: $\beta_4$

\includegraphics[width=0.9\linewidth]{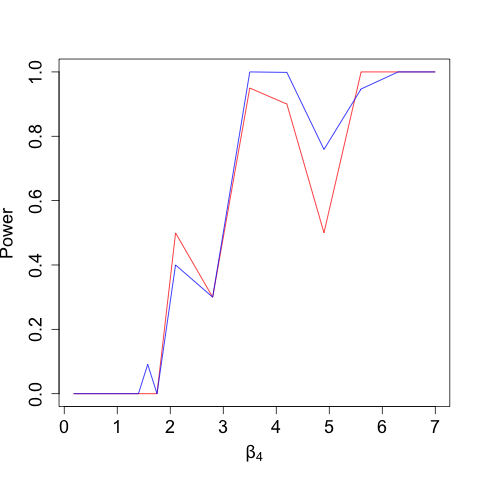}

3) Population size: $d$

\includegraphics[width=0.9\linewidth]{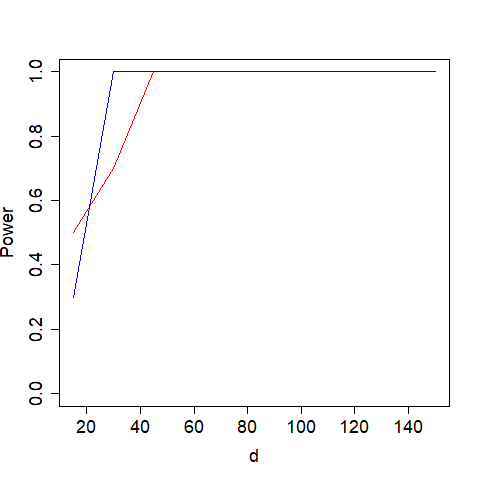}
\end{minipage}
\begin{minipage}[t]{0.48\linewidth}

2) Sample size: $n$

\includegraphics[width=0.9\linewidth]{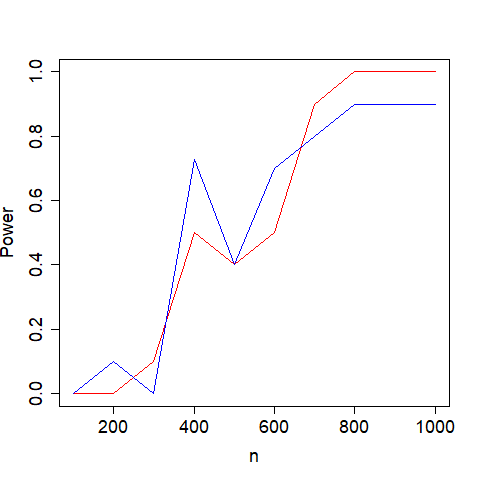}

4) Correlation of misspecified leave: $r$

\includegraphics[width=0.9\linewidth]{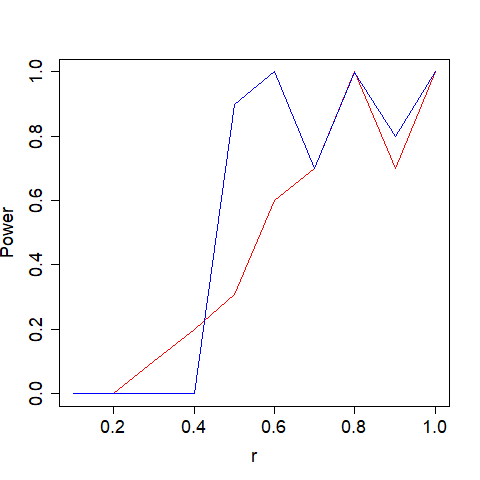}
\end{minipage}
\caption{Dependence of power to detect $L_4$ for Jeffreys prior (red) and the robust g-prior (blue) when varying different parameters as specified above each plot. Parameters which are not explicitly varied are kept fixed at the levels from the original Scenario 5, except for the first plot where all four coefficients $\beta_1 \dots, \beta_4$ are simultaneously varied by multiplying with the same factor. }\label{Fig:sens_beta4}
\end{figure}

We performed sensitivity analysis 
for the power to detect trees in Scenario 5 based on $\tilde{P}(L_j|Y)>0.5$ for $j \in \{1, \dots, 4\}$. Figure \ref{Fig:sens_beta4} presents the results for the four-way interaction $L_4$. Results for the trees with fewer leaves are provided in Figures S1 - S3 of the web supplement. Specifically we wanted to study how  the power is effected by the following factors: 
\begin{enumerate}
\item A change in the corresponding \textbf{coefficients} $\beta_j$, where all coefficients are varied simultaneously by multiplying them with a factor $K \in \{0.05,0.1,0.2,\dots, 1\}$ and all other parameters are kept constant.
\item A change in the \textbf{sample size} $n$, where the sample size $n$ is varied from $100$ to $1000$ and all other parameters are kept constant.

\item A change in the \textbf{population size} $d$, where the population size $d$ is varied from $15$ to $150$ and all other parameters are kept constant.
\item A \textbf{misspecified leave} within $L_4$, where the misspecified leave is substituted by a correlated leave with the correlation $r$ varying from 0.1 to 1. 
\end{enumerate}
In cases 2, 3 and 4 the relevant parameters were increased uniformly in 10 steps, in all cases $k_{max}$ was set to $20$. 
For computational reasons the sensitivity analysis was performed only using 10 simulation runs for each parameter value, both for Jeffreys prior and for the robust g-prior.  This number of repetitions is not sufficient to give high resolution estimates of the power but it is enough to illustrate the general dependence on each of the considered parameters.

The first two plots of Figure~\ref{Fig:sens_beta4} illustrate how the power to detect $L_4$ changes when either the regression coefficient $\beta_4$ or the sample size $n$ are varied. With a sample size of 1000 the power seems to deteriorate only for effect sizes smaller than 4, whereas for the large effect size of Scenario 5  a sample size of $n = 600$ still seems to provide reasonable power to detect $L_4$. The first plots of Figures S1 - S3 of the web supplement show that for the lower order trees a sample size of $n = 1000$ is sufficient to obtain reasonable power for much smaller effect sizes. Notably the three-way interaction $L_3$ can be detected with large power already for $\beta_3 = 1$ which is of the same order as the standard deviation of the error term. 

To reach sufficient power to detect four-way interactions with smaller regression coefficients one would have to increase the sample size. For many statistical models there is the notion that when decreasing the effect size by a factor $1/K$ one would roughly have to increase the sample size by a factor $K^2$ to end up with the same power. Figure S4 from the web supplement indicates that this relationship also holds for the logic regression approach and together with the results from the first plot of Figure~\ref{Fig:sens_beta4} one can induce that a sample size of $n > 10000$ is needed to have sufficient power to detect four-way interactions with regression coefficients which are of the order of the error standard deviation.

The third plot of Figure~\ref{Fig:sens_beta4} is concerned with the influence of the population size $d$ from the GMJMCMC algorithm on the power to detect $L_4$. Corresponding plots for the trees of lower size, for which the power is almost always equal to one, are provided in the web-supplement.  As one can see for both  priors power to detect $L_4$ grows gradually from $0$ to $1$ when $d$ changes from $15$ to $45$.
For  values of $d>30$ the power remains stable at 1. This illustrates the statement of Theorem~\ref{th:GMJMCMC}, according to which one requires $d-d_1\geq k_{max}$ to have an irreducible algorithm in the restricted space of logic regression models. In these simulations we have $k_{max} = 20$ and $d_1 = 10$. Hence according to Theorem \ref{th:GMJMCMC} a population size $d \geq 30$ is  sufficient for asymptotic irreducibility of the GMJMCMC algorithm.  For $d-d_1 < k_{max}$ irreducibility is no longer guaranteed and hence we cannot expect the approximations of the model posteriors to be precise  in all cases, specifically when the model size of the data generating model is larger than $d-d_1$.

The final plot of Figure~\ref{Fig:sens_beta4} considers the effect of misspecification of one leave. This setting is motivated by genetic association studies, where it often happens that not a causal SNP itself is genotyped but rather a strongly correlated tag SNP. As long as the correlation of the misspecified leave to the original leave is larger than 0.5 there appears to be no dramatic loss of power which indicates that a certain amount of model misspecification can be tolerated by our method.

\subsection{Analysis of Arabidopsis data}

According to our simulation results there is no large difference in the performance of GMJMCMC between using Jeffreys prior or the robust g-prior. On the other hand the clear computational advantage of Jeffreys prior seems to justify to omit the robust g-prior for analyzing real data. Hence in this section we only use Jeffreys prior for GMJMCMC. Furthermore we used $k_{max} = 15$ and $d = 25$ which allows for way more complex models than we would expect to see.  

 \cite{Balasub} mapped several different quantitative traits in \textit{Arabidopsis thaliana} using an advanced intercross-recombinant inbred line (RIL). Their data is publicly available as supporting information of their PLOS ONE article \citep{Balasub} which also gives all the details of the breeding scheme and the measurement of the different traits. We consider here only the hypocytol length in \textit{mm} under different light conditions \footnote{Data obtained from the second to fifth column of the file \url{
http://journals.plos.org/plosone/article/file?type=supplementary&id=info:doi/10.1371/journal.pone.0004318.s002}}. 
 Genotype data is available for 220 markers distributed over the 5 chromosomes of Arabidopsis thaliana with 61, 39, 43, 31 and 46 markers, respectively. \cite{Balasub} had genotyped 224 markers  
but we dismissed 4 markers which had identical genotypes with other markers.   The amount of missing genotype data is relatively small with a genotype rate of ~93.9\% and most importantly the data contains only homozygotes  (AA:49.6\% vs. BB:50.4\%). This means that the RIL population contains no heterozygote markers and logic regression can be directly applied using the genotype data as Boolean variables. Missing data were imputed using the R-QTL package (\url{http://www.rqtl.org/}).

The imputed data was then analyzed with our algorithm GMJMCMC to detect potential epistatic effects and the results are summarized in Table \ref{RDA}. Under blue light~\cite{Balasub} reported 4 potential QTL's, the strongest one on chromosome 4 in the regions of marker X44606688
 and three further fairly weak QTL on chromosomes 2, 3 and 5. Our analysis based on logic regression confirmed X44606688 and also detected those markers on chromosomes 2 and 5, though with a posterior probability slightly below 0.5. There was also some indication of a two-way interaction between the strong QTL on chromosome 4 and the QTL on chromosome 2.

\begin{table}
\centering

\begin{tabular}{|l|l|l|c|l|}\hline
Phenotype&Chr&Marker expression&$\tilde P(L\mid Y)$&Signif.\\\hline
Blue Light&4&X44606688 &   0.767&***\\
Blue Light&5&X44607250 &   0.335&**\\
Blue Light&2&X21607656 &   0.309&**\\
Blue Light&4$\wedge$2&X44606688$\wedge$X44606810 & 0.203&*\\\hline
Red Light&2&MSAT2.36&  0.441&**\\
Red Light&2&PHYB&  0.353&**\\
Red Light&2$\wedge$1&PHYB$^c\wedge$X44606541&  0.112&*\\
Red Light&2&X21607013&     0.092&*\\\hline
Far Red Light&4&MSAT4.37&  0.302&**\\
Far Red Light&4&NGA1107&   0.302&**\\\hline
White Light&5&X44606159&   0.632&***\\
White Light&1&X21607165&   0.427&**
\\\hline
\end{tabular}
\caption{\label{RDA}Potential additive and epistatic QTL for hypocytol length under different light conditions for Arabidopsis thaliana. Recombinant inbreed line data set taken from~\cite{Balasub}. The last column shows the level of confidence with *** corresponding to $\tilde P(L\mid Y)>0.5$, ** to $\tilde P(L\mid Y)>0.3$ and $*$ to $\tilde P(L\mid Y)>0.05$.} 
\end{table}

Under red light the original interval mapping analysis reported the region of MSAT2.36 as a strong QTL on chromosome 2 and x44607889
as a weaker QTL on chromosome 1. Our logic regression analysis distributes the marker posterior weights on three different markers  on chromosome 2 which are all in the neighborhood of MSAT2.36. Additionally there is some rather small posterior probability for an epistatic effect between this region and a marker on chromosome 1 which is rather close to x44607889. 
Finally both for Far Red Light and for White Light our analysis essentially yielded the same results as the interval mapping analysis, when observing that under the first condition the posterior probability was again almost equally distributed between the neighboring markers MSAT4.37 and NGA1107.  
In summary  the sample size in this data set might be slightly too small to detect epistatic effects, although under the first two light conditions there was at least some indication for a two-way interaction.

We have analyzed a second data set concerned with QTL mapping for Drosophila where we compare logic regression with a more traditional approach to modeling epistasis. Further details and results are presented in Section D of the web supplement.

\section{Discussion}

We have introduced GMJMCMC as a novel algorithm to perform Bayesian logic regression and compared it with the two existing methods MCLR \citep{Koop} and FBLR \citep{Fritsch2}. The main advantage of GMJMCMC is that it is designed to identify more complex logic expressions than its predecessors. Our approach differs both in terms of prior assumptions and in algorithmic details. Concerning the prior of regression coefficients we compared the simple Jeffreys prior with the robust g-prior. Jeffreys prior in combination with the Laplace approximation coincides with a BIC-like approximation of the marginal likelihood, which was also used by MCLR. The robust g-prior has some very appealing theoretical properties for the linear model. However, in our simulation study it gave only slightly better results than Jeffreys prior for the linear model and in case of logistic regression actually  performed worse in terms of power to detect the trees of the data generating logic regression model. 
With respect to the model topology we chose a prior which is rather similar to the one suggested by \cite{Fritsch2} for FBLR, but instead of using a truncated geometric prior for the number of leaves of a tree we suggest a prior which penalizes the complexity of a tree indirectly proportionally to the total number of trees of a given size. The motivation behind this prior is to control the number of false positive detections of trees in a similar way to how the Bonferroni correction works in multiple testing.

GMJMCMC has the capacity to explore a much larger model search space than MCLR and FBLR because it manages to efficiently resolve the issue of not getting stuck in local extrema, a problem that both MCLR and FBLR have in common. In logic regression the marginal posterior probability function is typically multi-modal in the space of models, with a large number of extrema which are often rather sparsely located. Additionally, the search space for logic regression is extremely large, where even computing the total number of models is a sophisticated task. As discussed in more detail in \cite{hubin2016efficient}, in such a setting simple MCMC algorithms  often get stuck in local extrema, which significantly slows down their performance and convergence might only be reached after run times which are infeasible in practice.

The success of GMJMCMC relies upon resolving the local extrema issue, which is mainly achieved by combining the following two ideas. First, when iterating through a fixed search space $S$, GMJMCMC utilizes the MJMCMC algorithm \citep{hubin2016efficient} which was specifically constructed to explore multi-modal regression spaces efficiently. Second, the evolution of the search spaces is governed within the framework of a genetic algorithm where a population consists of a finite number of trees forming the current search space. The population is updated by discarding trees with low estimated marginal posterior probability and generating new trees with a probability depending on the approximations of marginal inclusion probabilities from the current search space. The aim of the genetic algorithm is to converge towards a population which includes the most important trees.
Finally the performance of GMJMCMC is additionally boosted by running it in parallel with different starting points. 
 Irreducibility of the proposals both for search spaces and for models within the search spaces guarantees that asymptotically the whole model space will be explored by GMJMCMC and global extrema will at some point be reached under some weak regularity conditions. Clearly the genetic algorithm used to update search spaces results in a Markov chain of model spaces.

One important question in the context of logic regression is concerned with how to define true positive and false positive detections in simulations. We adapted a rather strict point of view which might be called an 'exact tree approach': Only those detected logic expressions which were logically equivalent with  trees from the data generating model were counted as true positives. While this seems to be a natural definition there are  certain pitfalls and ambiguities that occur in logic regressions which might speak against this strict definition.
 Apart from the more obvious logic equivalences according to Boolean algebra, for example due to De Morgan's laws or the distributive law, there can be slightly more hidden logic identities in logic regression. For example the expressions $(X_1  \vee X_2) - X_1$ and $X_2 - (X_1  \wedge X_2)$ give identical models. We have seen a less trivial example including four-way interactions in Scenario 6 of our simulation study, where the data generating tree $L_8$ is equivalent to the expression $X_{11}\wedge X_{13} + X_{19}\wedge X_{50} - X_{11}\wedge X_{13}\wedge X_{19}\wedge X_{50}$ consisting of three trees. 
Furthermore, different logic expressions can be highly correlated even when they are not exactly identical. 

Especially the results from the most complex Scenario~6 impose the question whether the exact tree approach is slightly too strict to define false positives. Subtrees of true trees give valuable information even if they are not describing the exact interaction. Often combinations of several subtrees and trees with misspecified logical operators can give expressions which are very close to the correct interaction term. For Scenario 6 we reported two possible summaries of the simulation results, one based strictly on the exact tree approach and the other one counting simultaneous detections of $X_{11}\wedge X_{13}, X_{19}\wedge X_{50}$ and $X_{11}\wedge X_{13}\wedge X_{19}\wedge X_{50}$ also as true positives. This was slightly ad hoc and we believe that good reporting of logic regression results is an area which needs further research.  
The output of MCLR takes a step in that direction, where only the leaves of trees are reported and if a tree has been detected then also all its subtrees are reported. However, in our opinion MCLR throws away too much information. 
We believe that several different layers of reporting might be more desirable, for example the exact tree approach, the MCLR approach and then something in between which does not reduce trees completely to their set of leaves. We have started to think more systematically in that direction and leave this topic open for another publication.

%Our simulation study demonstrated the potential of the GMJMCMC algorithm to find true logical expressions with high power and low false discovery rate, whilst in the real data examples GMJMCMC could find interesting epistatic effects in QTL analysis. However, the current implementation has a slight tendency to prefer a set of several simple trees over a single complicated tree. Specifically it does not properly take into account that a complex tree can be represented in several equivalent ways which leaves space for further improvements.

This paper has had a focus on model selection and selection of features of interest. The method is however directly applicable to prediction as well. %Standard Bayesian model averaging can then easily be applied.
One can approximate the posterior probability of some parameter/variable $\Delta$ via model averaging by
\begin{equation*}\label{Approx_post_delta}
\tilde P(\Delta\mid Y) = \sum_{M\in \Omega^*} P(\Delta\mid M,Y)\tilde P(M\mid Y)\;,
\end{equation*}
where $\Delta$ might be for example the predictor of unobserved data based on a specific set of covariates. 
Given estimates of posterior model probabilities, other prediction procedures such as the median probability model~\citep{barbieri2004optimal} or the posterior weighted median~\citep{clarke2013prediction} can also easily be applied.

\begin{supplement}
\sname{Supplementary materials}\label{suppA} 
%\stitle{Supplementary materials}
\slink[url]{see ReadMe file for details} 

\url{https://github.com/aliaksah/EMJMCMC2016/tree/master/supplementaries/Bayesian\%20Logic\%20Regression}
\sdescription{\\The file WebSupplement.pdf (\url{doi:10.1214/18-BA1141SUPP}) is provided as  supplementary material for this paper.}
\end{supplement}

%\begin{acknowledgement}
%The first two authors gratefully acknowledge the financial support of the \textit{CELS project at the %University of Oslo}, \url{http://www.mn.uio.no/math/english/research/groups/cels/index.html}. 
%\end{acknowledgement}

\bibliographystyle{ba}
\bibliography{sample}

\end{document}